\documentclass[11pt,a4paper,english,superscriptaddress]{revtex4}
\usepackage{amssymb}
\usepackage{amsmath} 
\usepackage{slashed}
\usepackage{graphicx}
\usepackage{babel}
\usepackage{hyperref}
\usepackage{color}
\usepackage{comment}
\makeatletter\AtBeginDocument{\let\@elt\relax}\makeatother

\begin{document}

\title{Gravitational corrections to the two-loop beta function in a non-Abelian gauge theory}

\author{M. Gomes} \email{mgomes@fma.if.usp.br} \affiliation{Instituto de F\'\i sica, Universidade de S\~ao Paulo, Caixa Postal 66318, 05315-970, S\~ao Paulo, S\~ao Paulo, Brazil.}

\author{A.~C.~Lehum} \email{lehum@ufpa.br} \affiliation{Faculdade de F\'isica, Universidade Federal do Par\'a, 66075-110, Bel\'em, Par\'a, Brazil.}

\author{A. J. da Silva} \email{ajsilva@fma.if.usp.br} \affiliation{Instituto de F\'\i sica, Universidade de S\~ao Paulo, Caixa Postal 66318, 05315-970, S\~ao Paulo, S\~ao Paulo, Brazil.}
	
\begin{abstract}
This paper investigates the coupling of massive fermions to gravity within the context of a non-Abelian gauge theory, utilizing the effective field theory framework for quantum gravity. Specifically, we calculate the two-loop beta function of the gauge coupling constant in a non-Abelian gauge theory, employing the one-graviton exchange approximation. Our findings reveal that gravitational corrections may lead to a non-trivial UV fixed point in the beta function of the gauge coupling constant, contingent upon the specific gauge group and the quantity of fermions involved. 
\end{abstract}

\maketitle

\section{Introduction}

One of the most important properties of non-Abelian gauge theories is the asymptotic freedom. Discovered by David Gross and Frank Wilczek \cite{Gross:1973id} and independently by David Politzer \cite{Politzer:1973fx} in 1973, this phenomenon entails a decrease in the strength of the gauge coupling constant as energy scales increase. This attribute assumes critical significance whenever employing non-Abelian gauge theories for the representation of strong interactions~\cite{Gross:1974cs}.

In spite of the non-renormalizability inherent in Einstein's theory of gravity when quantized for small fluctuations around a flat metric \cite{'tHooft:1974bx, PhysRevLett.32.245,Deser:1974cy}, Robinson and Wilczek, in 2005, employed the effective field theory approach to quantum gravity~\cite{Donoghue:1994dn} to address the issue of how gravity might impact the asymptotic behavior of gauge theories \cite{Robinson:2005fj}. They suggested that gravitational corrections lead to the asymptotic freedom of the  gauge coupling constants. Due to the dimensional nature of the gravitational coupling constant $\kappa=\sqrt{32\pi G}=\sqrt{32\pi}/M_P$, where $G$ is the Newton's constant, and $M_{P}$ is the Planck's mass, this proposition was grounded in the emergence of the dimensionless combination $\kappa^2 E^2$, where the UV cutoff $E$ was interpreted as an energy scale, arising from quadratic UV divergent Feynman diagrams.

However, Pietrykowski later contested this conclusion, demonstrating its gauge dependence \cite{Pietrykowski:2006xy}. Subsequently, numerous studies have been conducted to explore the application of the renormalization group in quantum gravity as an effective field theory~\cite{Felipe:2012vq,Ebert:2007gf,Nielsen:2012fm,Toms:2008dq,Toms:2010vy,Ellis:2010rw,Anber:2010uj,Folkerts:2011jz,Bevilaqua:2015hma,Bevilaqua:2021uzk, Bevilaqua:2021uev}. In particular, the authors of Ref.\cite{Folkerts:2011jz} determined that under the condition of preserving all symmetries, in the Yang-Mills theory, the weak gravity limit does not receive any contributions from the gravitational sector to the running gauge coupling.

Different types of dimensionless combinations can arise within Feynman amplitudes, including $\kappa^2 p^2$ and $\kappa^2 m^2$ when massive particles are involved in internal loops. The combination $\kappa^2 p^2$ plays a pivotal role in characterizing the renormalization of high-order operators within the framework of the effective field theory approach, while the $\kappa^2 m^2$ combination is pertinent to the renormalization of marginal operators. Therefore, the presence of massive particles becomes a crucial factor. Previous studies \cite{Toms:2008dq,Toms:2009zz,Toms:2009vd} have explored the influence of another dimensionful parameter, the cosmological constant, which manifests in gravitational corrections to the gauge coupling beta function through the dimensionless combination $\kappa^2\Lambda$. However, in the present work, we opt not to address the cosmological constant, maintaining a flat background metric.

Numerous studies have concluded that gravitational corrections to the beta function of the gauge coupling constant are absent at the one-loop order (see, for example, \cite{Ebert:2007gf,Nielsen:2012fm,Toms:2008dq,Charneski:2013zja,Bevilaqua:2015hma,Bevilaqua:2021uev}; see also \cite{Comment}). However, two-loop corrections have been examined in \cite{Bevilaqua:2021uzk}, revealing gravitational corrections to the beta function of the electric charge in Quantum Electrodynamics (QED) at this order. Despite these gravitational corrections, the electric charge does not exhibit asymptotic freedom and lacks a nontrivial fixed point. Indeed, these corrections contribute positively to the beta function, as expressed by the equation
\begin{equation}\label{betaQED}
	\beta(e) = \frac{e^3}{12\pi^2} + \frac{e^5}{128\pi^4} + \frac{5e^3m^2}{24\pi M_P^2},
\end{equation}
where $m$ is the mass of the fermion (matter) field. However, since gravity is universally attractive, it is anticipated that in the non-Abelian case, the gravitational contribution to the beta function should be also positive. This has the potential to undermine asymptotic freedom, particularly in proposed extensions of the Standard Model involving fermions (or scalars) with masses approaching the order of the Planck mass.

With this concern in mind, the objective of our study is to calculate the two-loop beta function of the gauge coupling constant within the framework of a non-Abelian gauge theory, considering the one-graviton exchange approximation. Our findings reveal that the beta function of the gauge coupling constant exhibits a non-trivial UV fixed point, proportionally related to the ratio $m^2/M_P^2$. In the Standard Model (SM), this ratio is expected to be associated with the mass of the top quark, expressed as $m_t^2/M_P^2\sim 10^{-34}$. However, when contemplating proposals for extensions of the SM involving fermions with masses approaching the order of the Planck mass, gravitational contributions may become more significant.

The structure of this paper is organized as follows: In Section \ref{sec2}, we provide an introduction to the model. Section \ref{subsec3.1} is dedicated to presenting general arguments regarding the calculation of gravitational corrections to the two-loop beta function of the gauge coupling constant. The computation of gravitational corrections to the two-loop gauge field self-energy is carried out in Section \ref{subsec3.2}, facilitating the determination of the gravitational correction to the two-loop beta function of the gauge coupling constant in Section \ref{subsec3.3}. Concluding remarks are presented in Section \ref{final}. Throughout this study, the Minimal Subtraction (MS) scheme is employed to handle divergences, and natural units with $c=\hbar=1$ are used.

\section{The EFT for a non-Abelian gauge theory with fermions coupled to gravity}\label{sec2}

We initiate our investigation with the Lagrangian outling the effective field theory (EFT) for a non-Abelian gauge theory, incorporating fermions interacting with gravity:
\begin{eqnarray}\label{fQCD} 
	\mathcal{L}=&& \sqrt{-g}\sum_f\Big\{\frac{2}{\kappa^2}R-\frac{1}{4} g^{\mu\rho}g^{\nu\sigma} G_{\mu\nu}^{ a} G_{\rho\sigma}^a +i\bar{\psi}_f(\nabla_{\mu} - igA_{\mu}^at^a)\gamma^{\mu}\psi_f - m_f\bar{\psi}_f\psi_f \Big\}, 
\end{eqnarray}

\noindent where the index $f=1,2,\cdots,N_f$ spans over the fermion flavors, and $G^a_{\mu\nu}=\partial_\mu A_\nu^a-\partial_\nu A_\mu^a + gf^{abc}A^b_\mu A^c_\nu$ denotes the non-Abelian field-strength, with $t^a$ being the SU(N) generators and $f^{abc}$ representing the structure constants of the $SU(N)$ group. The Dirac matrices are contracted with the vierbein ($\gamma^{\mu} \equiv \gamma^\alpha e^{\mu}_\alpha$, $g_{\mu\nu}=e^{\alpha}_{\mu} e^{\beta}_{\nu} \eta_{\alpha\beta}$, $\overrightarrow{\nabla}_{\mu}\psi=(\partial_\mu +i\omega_\mu)\psi$, $\bar\psi\overleftarrow{\nabla}_{\mu}=(\partial_\mu\bar\psi -i\bar\psi\omega_\mu)$, $\omega_\mu=\frac{1}{4}\sigma^{\alpha\beta}\left[e^\nu_\alpha(\partial_\mu e_{\beta\nu}-\partial_\nu e_{\beta\mu})
+\frac{1}{2}  e^\rho_\alpha e^\sigma_\beta(\partial_\sigma e_{\gamma\rho}-\partial_\rho e_{\gamma\sigma})e^\gamma_\mu-(\alpha\leftrightarrow \beta)\right]$ is the spin connection with $\sigma^{\alpha\beta}=i[\gamma^\alpha , \gamma^\beta]/2$). Here we use greek letters from the middle and the beginning of the alphabet to denote general and  locally inertial coordinates, respectively. 

To conform to the effective field theory framework of gravity, it becomes imperative to expand $g_{\mu\nu}$ around the flat metric as detailed below:
\begin{equation}\label{metric} 	
	g_{\mu\nu} = \eta_{\mu\nu} + \kappa h_{\mu\nu} \quad (\text{exactly}), \quad g^{\mu\nu} = \eta^{\mu\nu} - \kappa h^{\mu\nu} +\cdots, 
\end{equation}
where the spacetime indices (Greek) are raised and lowered utilizing the flat metric $\eta_{\mu\nu}=(+,-,-,-)$. As we are restricting ourselves to the one-graviton exchange approximation, the basic parts of the effective Lagrangian ${\cal L}$ are:
\begin{eqnarray}
{\cal L}= {\cal L}_{h}^{0}+ {\cal L}_{f}+ {\cal L}_{A}, 
\end{eqnarray}
{ where
\begin{eqnarray}
	\mathcal{L}_{h}^{0} &=& 
	\frac{1}{2}\partial^{\rho}h_{\mu\nu} \partial_{\rho}h^{\mu\nu}
	-\frac{1}{2}\partial^\mu h \partial_\mu h
	-\partial^{\mu}h_{\mu\nu}\partial_{\rho}h^{\rho\nu}
	+\partial^{\mu}h_{\mu\nu}\partial_{\nu}h,
\end{eqnarray}
with $h=h^{\mu}_{\mu}$, is the Lagrangian for the gravitational sector without self-interaction terms,
\begin{subequations}
\begin{eqnarray}
  \mathcal{L}_f &=& \mathcal{L}_f^0 + g\bar{\psi}_f\gamma^\mu A_\mu^a t^a\psi_f  + \kappa\mathcal{L}_f^1 + \cdots;\\
  \mathcal{L}_f^0 &=& \frac{i}{2}(\bar{\psi}_f\gamma^\mu\partial_\mu\psi_f - \partial_\mu\bar{\psi}_f\gamma^\mu\psi_f) - m_f\bar{\psi}_f\psi_f;\\
  \mathcal{L}_f^1 &=& \frac{1}{2}h\mathcal{L}_f^0 - \frac{i}{4}h_{\mu\nu} (\bar{\psi}_f\gamma^\mu\partial^\nu\psi_f- \partial^\nu\bar{\psi}_f\gamma^\mu\psi),
\end{eqnarray}
\end{subequations}
for the fermion sector, and 
\begin{subequations}\label{LA}
 \begin{eqnarray}
  \mathcal{L}_A &=& \mathcal{L}_A^0 + \kappa\mathcal{L}_A^1 + \cdots ;\\
  \mathcal{L}_A^0 &=& -\frac{1}{4}G_{\mu\nu}^aG^{\mu\nu}_a; \\
  \mathcal{L}_A^1 &=& \frac{1}{2}h^\tau_{~\nu}G^{\mu\nu}_aG_{\mu\tau}^a + \frac{1}{2}h\mathcal{L}_A^0,
 \end{eqnarray}
\end{subequations}
for the gauge sector.
A detailed expansion of the Lagrangian given by equation \eqref{fQCD} is provided in Ref. \cite{Choi:1994ax}.

We now proceed with the quantization of the model, following the Faddeev-Popov procedure. This involves introducing the gauge-fixing and ghost fields for both the vector and tensor fields. The gauge-fixing Lagrangian is expressed as
\begin{eqnarray}
	\mathcal{L}_{GF}= \frac{1}{2\xi_A}(\partial^\mu A_\mu)^2 + \frac{1}{2\xi_h}\left(\partial_\mu h^{\mu\nu}-\frac{1}{2}\partial^\nu h \right)^2.
\end{eqnarray}
Given that we are operating within the one-graviton exchange approximation, there is no necessity to explicitly include the ghosts for the graviton.  The gauge ghosts Lagrangian is given by
\begin{equation}\label{ghost}
	\mathcal{L}_{ghost} = \sqrt{-g}~g^{\mu\nu}\partial_\mu\bar{c}^a(\partial_{\nu}\delta^{ac} + g f^{abc}A^b_\nu)c^c = \partial^{\mu}\bar{c}^a(\partial_{\mu}\delta^{ac} + g f^{abc}A^b_\mu)c^c+\mathcal{O}(\kappa).
\end{equation}
As the gauge ghost turns out to be massless its contribution in our calculations will be always "higher order" $\kappa^2 p^2$ and therefore innocuous to our results; in what follows they will be omitted.

To render the model "renormalizable" up to order $p^2/M_P^2$ in the momentum expansion, enabling it to absorb all potential divergences arising in the perturbative expansion up to order $p^2/M_P^2$ (where $p$ denotes external momenta), it becomes necessary to incorporate a Lagrangian comprising higher derivative terms. The pertinent terms for our objectives are 
\begin{eqnarray}\label{ho}
\mathcal{L}_{HO} &=& i\bar\psi_f~\frac{\Box}{M_P^2}\left(\tilde{g}_1\slashed{\partial}- \tilde{g}_2 m_f\right)\psi_f-\frac{\tilde{g}_3}{4M_P^2}G^{\mu\nu}_a\Box G_{\mu\nu}^a +\frac{i\tilde{g}_4}{2M_P^2}\bar\psi_f t^a\gamma_\mu\partial_\nu\psi_f G^{\mu\nu}_a 
+ \cdots,
\end{eqnarray}
\noindent where $\tilde{g}_i$ denote dimensionless coupling constants. {
 From the above expressions, we may obtain the propagators of the model which  are the usual ones, given by
\begin{subequations}\label{propagators}
\begin{eqnarray}
S_F(p) &=& i\frac{\slashed{p}+m_f}{p^2-m_f^2};\\
\Delta^{\mu\nu}_{ab}(p) &=& \frac{i}{p^2}\left(\eta^{\mu\nu}-(1-\xi_A)\frac{p^\mu p^\nu}{p^2} \right)\delta_{ab};\\
\Delta_{ab}(p) &=& \frac{i}{p^2}\delta_{ab};\\ 
\Delta^{\rho\sigma\mu\nu}(p) &=& \frac{i}{p^2}\left(P^{\rho\sigma\mu\nu}-(1-\xi_h)\frac{Q^{\rho\sigma\mu\nu}}{p^2}\right),  
\end{eqnarray}
\end{subequations}
\noindent where $S_F(p)$, $\Delta^{\mu\nu}_{ab}(p)$, $\Delta_{ab}(p)$ and $\Delta^{\rho\sigma\mu\nu}(p)$ represent the propagators for fermions, gluons, ghosts, and gravitons, respectively.  The projectors $P^{\rho\sigma\mu\nu}$ and $Q^{\rho\sigma\mu\nu}$  are given by
\begin{eqnarray}
P^{\rho\sigma\mu\nu} &=&\frac{1}{2} \left(\eta^{\rho\mu}\eta^{\sigma\nu}+\eta^{\rho\nu}\eta^{\sigma\mu}-\eta^{\rho\sigma}\eta^{\mu\nu} \right);\nonumber\\
Q^{\rho\sigma\mu\nu} &=& (\eta^{\rho\mu}p^\sigma p^\nu+\eta^{\rho\nu}p^\sigma p^\mu+\eta^{\sigma\mu}p^\rho p^\nu+\eta^{\sigma\nu}p^\rho p^\mu).
\end{eqnarray}

To investigate the renormalization of the model, we initiate by redefining the fields and parameters in the Lagrangian \eqref{fQCD}. For instance, the vector and fermion  field strengths are redefined as $A^\mu_a\rightarrow Z_3^{1/2}A^\mu_a$ and $\psi_f\rightarrow Z_{2f}^{1/2}\psi_f$, where $Z_i$ represent the renormalizing functions, structured as a perturbative series in the number of loops, given by
\begin{eqnarray}\label{ctdelta}
Z_i=Z_i^{(0)}+ Z_i^{(1)}+Z_i^{(2)}+\cdots=1+\delta_i, \qquad \text{with} \quad Z_i^{(0)} = 1.
\end{eqnarray}

The relationship between the bare ($g_0$) and the renormalized ($g$) gauge coupling constants in terms of the $Z$ functions can be expressed in four distinct ways: 
\begin{subequations}\label{g0}
\begin{eqnarray}\label{eq_e_0}
g&=&\mu^{-2\epsilon}\frac{Z_2 Z_3^{1/2}}{Z_1}g_0;\label{g0Z1Z2Z3}\\
g&=&\mu^{-2\epsilon}\frac{Z_{3}^{3/2}}{Z_{3g}}g_0;\label{g0Z3g}\\
g&=&\mu^{-2\epsilon}\frac{Z_{3}}{Z_{4g}^{1/2}}g_0;\label{g0Z4}\\
g&=&\mu^{-2\epsilon}\frac{Z_{2c}Z_3^{1/2}}{Z_{1c}}g_0,
\end{eqnarray}
\end{subequations}

\noindent where $\mu$ represents a mass scale introduced by dimensional regularization (DR) to regulate the UV divergences in the Feynman amplitudes, $\epsilon$ is associated with the spacetime dimension $D$ via $D=4-2\epsilon$, $Z_1=(1+\delta_1)$ denotes the gauge coupling constant counterterm, $Z_{3g}=(1+\delta_{3g})$ signifies the counterterm that renormalizes the three-point function of the gluons, $Z_{4g}=(1+\delta_{4g})$ corresponds to the counterterm that renormalizes the four-point functions of the gluons, $Z_{2c}=(1+\delta_{2c})$ represents the wave function counterterm for the ghosts, and $Z_{1c}=(1+\delta_{1c})$ indicates the counterterm that renormalizes the gluon-ghost vertex.

In conjunction, Eqs. \eqref{g0} provide relations between the Green functions that must be satisfied to ensure gauge invariance, known as the Slavnov-Taylor identities. These relations, expressed in terms of the renormalizing functions $Z$, are summarized as
\begin{eqnarray}\label{Slavnov-Taylor}
 \frac{Z_1}{Z_2} &=& \frac{Z_{3g}}{Z_3}~ = \frac{Z_{4g}^{1/2}}{Z_3^{1/2}} = \frac{Z_{1c}}{Z_{2c}},
\end{eqnarray}
\noindent which in the MS procedure  can be perturbatively expressed  as
\begin{equation}\label{eqSTi01}
	\delta_1-\delta_2 = \delta_{3g} - \delta_3 = \frac{1}{2}\left(\delta_{4g} - \delta_3\right) = \delta_{1c} -\delta_{2c},
\end{equation}
\noindent where the counterterms $\delta_i$ are defined in \eqref{ctdelta}. 

In Ref. \cite{Souza:2022ovu}, the authors computed the counterterms at one-loop order in the presence of gravitational interaction, veryfying the validity of the Slavnov-Taylor identities ( As it is known, the aforementioned relations arise from the gauge invariance of the theory \cite{Slavnov:1972fg, Taylor:1971ff}). This implies that the formulation of a non-Abelian gauge theory (with or without gravity) results in the equations \eqref{g0}. Hence, we expect that they will hold true at any order in perturbation theory.

\section{Gravitational corrections to the two-loop gluon self-energy and beta function of the gauge coupling constant}\label{sec3}

\subsection{General argumentation}\label{subsec3.1}

Let us commence this section by outlining our approach to compute the gravitational corrections to the two-loop beta function of the gauge coupling constant. Utilizing Eq.\eqref{g0Z1Z2Z3}, we discern that the beta function of the gauge coupling constant $g$ can be determined via the relation
\begin{eqnarray}
	\beta(g) &=& \lim_{\epsilon\rightarrow0}\mu\frac{dg}{d\mu} =\lim_{\epsilon\rightarrow0}\mu\frac{d}{d\mu}\left[g_{0}\left(1-\delta_1+\delta_2+\frac{\delta_3}{2}\right)\mu^{-2\epsilon}\right].
\end{eqnarray}

Computing the gravitational corrections to the renormalization constants $Z_1^{(2)}$ and $Z_2^{(2)}$ can be arduous, involving approximately 100 diagrams. However, this laborious task can be circumvented by observing that from \eqref{eqSTi01}, $\delta_{1}-\delta_{2} = \delta_{1c}-\delta_{2c}$. Consequently,
\begin{eqnarray}
	\beta(g) &=& \lim_{\epsilon\rightarrow0}\mu\frac{dg}{d\mu} =\lim_{\epsilon\rightarrow0}\mu\frac{d}{d\mu}\left[g_{0}\left(1-\delta_{1c} +\delta_{2c} +\frac{\delta_3}{2}\right)\mu^{-2\epsilon}\right].
\end{eqnarray}

The renormalization constants $\delta_{1c}$ and $\delta_{2c}$ cannot depend on $\kappa$ because the corresponding functions of the gauge ghosts involve only massless particles within the closed loops, as evidenced by the self-energy process depicted in Figure \ref{figGhostSE01}. Thus, it is not possible to generate contributions proportional to $\kappa^2 m^2$. The gauge ghost self-energy is proportional to $\kappa^2p^4$, and the gauge ghost-gluon three-point function is proportional to $\kappa^2p^2p^\mu$ by similar arguments. Hence, their UV divergences must be absorbed by the renormalization of  high-order operators. Consequently, if $\delta_{1c}-\delta_{2c}$ is independent of $\kappa$, then $\delta_{1}-\delta_{2}$ must also be independent of $\kappa$. Therefore, the gravitational corrections to $\beta(g)$ must arise from the renormalization of the gluon self-energy, i.e., from the computation of $\delta_3$.

\subsection{Gravitational corrections to the two-loop gluon self-energy}\label{subsec3.2}

In this section, we focus on evaluating the gravitational corrections to the renormalization of the two-loop gluon self-energy. As discussed in the previous section, computing this function will suffice to determine the gravitational corrections to the two-loop beta function of the gauge coupling constant. The diagrams contributing to this process are illustrated in Figures \ref{figSE01}-\ref{figSE03}.

The diagrams depicted in Figure \ref{figSE01} represent conventional diagrams, reflecting scenarios without gravitational effects. Our calculations are consistent with existing literature \cite{Egorian:1978zx}. Figures \ref{figSE02} present gravitational corrections where no matter loops are involved. The collective contribution of these diagrams is necessarily proportional to $(p^2\eta^{\mu\nu}-p^\mu p^\nu)\kappa^2 p^2$, indicating their renormalization by the constant $\tilde{Z}_3$ associated with the high-order operator $G^{\mu\nu}_a\Box G_{\mu\nu}^a$.

Figures \ref{figSE03} depict gravitational corrections involving matter loops. The cumulative effect of these diagrams may contain terms proportional to $(p^2\eta^{\mu\nu}-p^\mu p^\nu)\kappa^2m^2$, suggesting their renormalization by the constant ${Z}_3$. Our current task is to compute these corrections.

To conduct this calculation, we constructed the amplitude using a suite of computational packages \cite{feynrules,Hahn:2000kx,Shtabovenko:2020gxv}. Although it is established that the full two-loop gluon self-energy must follow the form
\begin{eqnarray}\label{eq_pt_01}
	\Pi_{\mu\nu}^{ab}(p)=(p^2\eta_{\mu\nu}-p_\mu p_\nu)\Pi(p^2)\delta^{ab},
\end{eqnarray}
owing to gauge invariance and transversality of the gluon propagation, this expression isn't suitable as a benchmark during intermediate calculation steps since this property may not hold for individual diagrams.

Hence, our approach was to assume that, upon integration over internal momenta, each diagram $i$ should possess the more general (Lorentz invariant) form ${\Pi_i}_{\mu\nu}^{ab}(p)={\Pi_i}_{\mu\nu}\delta^{ab}=(\eta_{\mu\nu}p^2~A_i(p)+p_\mu p_\nu B_i(p))\delta^{ab}$, from which $A_i(p)$ and $B_i(p)$ can be derived through projections:
\begin{eqnarray} 	
	A_i &=& \frac{1}{(D-1)p^2}\left(\eta^{\mu\nu}-\frac{p^{\mu}p^\nu}{p^2}\right){\Pi_i}_{\mu\nu},\nonumber\\ 	B_i &=& -\frac{1}{(D-1)p^2}\left(\eta^{\mu\nu}-D\frac{p^{\mu}p^\nu}{p^2}\right){\Pi_i}_{\mu\nu}. \nonumber 
\end{eqnarray}
\noindent By summing over the diagrams, we arrived at the (anticipated) outcome $A(p)=-B(p)$, implying that the gluon polarization tensor adopts the transverse form of Eq. (\ref{eq_pt_01}). During these computations, we simplified the scalar two-loop integrals using the Tarasov algorithm \cite{Tarasov:1997kx}, aided by the computational package TARCER \cite{Mertig:1998vk} which reduces the calculation to some basic two-loop integrals which are available in \cite{TSIL}.

Finally, we evaluated the integrals  retaining only the ultraviolet (UV) divergent part of $\Pi(p^2=0)$, since these contributions are only logarithmically divergent. The detailed file containing these calculations can be found in the supplementary material \cite{site_lehum}, and the resulting expression for the UV divergent part of diagrams in Figure \ref{figSE03} is given by
\begin{eqnarray}
	-i\Pi_2(p)= -\frac{5 g^2 \kappa^2 M^2}{1536 \pi^4 \epsilon}
	+\mathcal{O}\left(p^2\right),
\end{eqnarray}
\noindent where $M^2$ represents the sum of the squared masses of the fermions. This term contributes to the gravitational correction in the renormalization of the $Z_3$ factor and consequently affects the two-loop correction to the beta function of $g$.

Furthermore, we need to calculate the one-loop diagrams with the insertion of the one-loop counterterms, whose amplitudes have the same order of  $g^2\kappa^2$ as the two-loop diagrams. These diagrams are depicted in Fig. \ref{figSECT}. The corresponding result has two parts. The first part, proportional to $g^4$, is related to the $QCD$ in the absence of gravity, and the corresponding result was studied in Ref. \cite{Caswell:1974gg}. The second part, the gravitational corrections of order $\mathcal{O}(g^2\kappa^2)$, is in fact proportional to $g^2\kappa^2 p^2$, corresponding to a gravitational correction to the higher-order $G^{\mu\nu}_a\Box G_{\mu\nu}^a$ operator, in such a way that no gravitational contribution to $G^{\mu\nu}_a G_{\mu\nu}^a$ comes from the diagrams depicted in Fig. \ref{figSECT}.
		
Hence, the gravitational correction to the renormalizing factor for the gauge field $Z_3$, at two-loop order, is given by 
\begin{eqnarray}\label{eqZ32l}
Z_3 &=& 1+\delta_3= 1 -\frac{5 g^2 \kappa ^2 M^2}{1536 \pi^4 \epsilon}+\cdots.
\end{eqnarray}

In the next section, we will utilize this result to calculate the gravitational correction to the beta function of the gauge coupling constant.

\subsection{Two-loop gauge coupling beta function}\label{subsec3.3}

With the inclusion of the gravitational corrections to the renormalization constant $Z_3$, as given in Eq. \eqref{eqZ32l}, the gravitational contribution up to order $\kappa^2$ is incorporated into the established result for the beta function of the non-Abelian gauge coupling constant, which has been documented in the literature \cite{Caswell:1974gg} as
\begin{eqnarray}
\beta(g)&=& -b_0\frac{g^3}{(16\pi^2)}+b_1\frac{g^5}{(16\pi^2)^2}+b_h\frac{g^3}{(16\pi^2)^2},
\end{eqnarray}
\noindent where $b_0=\left(\frac{11}{3}C_2(G)-\frac{4}{3}T(R)\right)$, $b_1=\left(-\frac{34}{3}C_2(G)^2+\frac{20}{3}C_2(G) T(R)+4 C_2(R)T(R)\right)$ and $\displaystyle{b_h=\frac{5\kappa^2M^2}{6}}$, with the group invariants $C_2(R)$, $C_{2}(G)$ being the quadratic Casimir operators for the fundamental and adjoint representations of the gauge group and $T(R)$ the trace for the fundamental representation. For QCD, where the gauge group is $SU(3)$, the coefficients are determined in the literature as $b_0 = (33 - 2N_f)/3$ and $b_1 = -2(153 - 19N_f)/3$.

It is convenient to define $\rho = g^2/4\pi$ and express $\beta(\rho)$ as
\begin{eqnarray}
	\beta(\rho)&=& -\frac{\rho^2}{2\pi}\left(b_0-\frac{b_1}{4\pi} \rho - \frac{b_h}{(4\pi)^2} \right)
	=-2\rho^2\left(b_0- \frac{b_1}{4\pi} \rho - \frac{5}{3\pi}\frac{M^2}{M_P^2}\right).
\end{eqnarray}
 
Notice that $\beta(\rho)$ can exhibit a nontrivial UV fixed point at
\begin{eqnarray}
	\rho_*=\frac{4\pi}{b1}\left(b_0-\frac{b_h}{(4\pi)^2} \right)=\frac{4\pi}{b1}\left(b_0-\frac{5}{3\pi}\frac{M^2}{M_P^2} \right).
\end{eqnarray}   

 For QCD, characterized by the gauge group $SU(3)$ and a fermion count of $N_f = 6$, the presence of a nontrivial UV fixed point $\rho_*$ is contingent upon $M^2 \geq 17 M_P^2$. Notably, $M^2$ in QCD approximates the squared mass of the top quark ($M_t = 172.76 \pm 0.3$ GeV), which is significantly smaller than the Planck mass ($M_P\approx 1.2\times 10^{19}$ GeV), thus precluding the existence of a nontrivial UV fixed point. If additional fermions are introduced by incorporating further quark generations (e.g. $N_f = 24$), the requirement for the existence of $\rho_*$ becomes $M^2 \geq 5.6 M_P^2$. In a very speculative situation (for example in the Kaluza Klein models, \cite{Gherghetta:2000qt,Burdman:2003ya}), 
 if the combined squared masses of the additional quarks exceed $5.6M_P^2$, a nontrivial UV fixed point for the strong coupling could emerge. 
This scenario could be applicable to other gauge groups and fermion counts as well.

\section{Final remarks}\label{final}

In summary, we have computed the gravitational corrections to the two-loop beta function of the gauge coupling constant within a non-Abelian gauge theory featuring fermions, utilizing the one-graviton exchange approximation. Our analysis demonstrates that the beta function of the gauge coupling constant experiences a positive gravitational correction at two-loop order, consistent with previous findings in Einstein-QED model\cite{Bevilaqua:2021uzk}. Furthermore, we observe that these gravitational corrections have the potential to generate a non-trivial UV fixed point in the beta function of the gauge coupling constant, dependent on the specific gauge group and the number of fermions present.

One important consideration arises from the critique presented in Ref. \cite{Anber:2010uj}. In their study, the authors argue that the physical evolution of coupling constants should be derived from S-matrix computations. Their findings suggest that incorporating gravitational effects into the evolution of couplings may not be universally applicable in describing physical phenomena. This discrepancy may stem from operator mixing between marginal and higher-order (irrelevant) operators in an on-shell renormalization process. Indeed, as discussed in Ref. \cite{Bevilaqua:2015hma}, which examined scattering processes in the Einstein-QED model, while operator mixing occurs between the $\lambda\phi^4$ operator and its higher-order counterpart (e.g., $\phi^3\Box\phi$), the same phenomenon does not affect the gauge coupling constant renormalization due to its specific kinematical dependence.  In our case we also found that there is no mixing directly involving the renormalization of the gauge coupling constant.

\acknowledgments
A. C. Lehum and A. J. da Silva are partially supported by Conselho Nacional de Desenvolvimento Cient\'ifico e Tecnol\'ogico (CNPq).  

\appendix
\section{The one-loop renormalization constants}\label{seC_2(G)ppendix}

In this appendix, we list the one-loop renormalization constants computed in Ref. \cite{Souza:2022ovu}, which are essential for our investigation. These constants are inserted as counterterm contributions in the one-loop diagrams depicted in Figure \ref{figSECT}. The one-loop renormalization constants are as follows:   
\begin{subequations}\label{ct01}
	\begin{eqnarray}
		\delta_2  &=& \frac{\kappa ^2 m_f^2 \left(29 \xi _h-37\right)}{512 \pi ^2 \epsilon }-\frac{C_2(R) g^2 \xi _A}{16 \pi ^2 \epsilon };\label{eq_z2f}\\
		\delta_{m^2} &=& \frac{\kappa ^2 m_f^2 \left(19 \xi _h-23\right)}{256 \pi ^2 \epsilon }-\frac{C_2(R) g^2 \left(\xi _A+3\right)}{16 \pi ^2 \epsilon };\label{ct01-2}\\
		\delta_3 &=& -\frac{g^2 \left(3 C_2(G) \xi _{A}-13 C_2(G)+4 N_f\right)}{96 \pi ^2 \epsilon };\label{eq_z3}\\
		\tilde\delta_3 &=& -\frac{\kappa^2(3\xi_h-2)}{96\pi^2\epsilon};\label{ct02-2}\\
		\delta_{2c} &=& -\frac{C_2(G) g^2 \left(\xi _A-3\right)}{64 \pi ^2 \epsilon };\\
		\delta_{1c} &=& -\frac{C_2(G) g^2\xi_A}{32 \pi ^2 \epsilon};\\
		\delta_{1} &=& \frac{\kappa ^2 m_f^2 \left(29 \xi _h-37\right)-8 g^2 \left(\xi _A \left(C_2(G)+4 C_2(R)\right)+3 C_2(G)\right)}{512 \pi ^2 \epsilon};\\
		\delta_{3g} &=& -\frac{g^2 \left(9 C_2(G) \xi _A-17 C_2(G)+8 N_f\right)}{192 \pi ^2 \epsilon};\\
		\delta_{4g} &=& -\frac{g^2 \left(3 C_2(G) \xi _A-2 C_2(G)+2 N_f\right)}{48 \pi ^2 \epsilon }.
	\end{eqnarray}
\end{subequations}

As evident from the equations above, the Slavnov-Taylor identities are upheld, as we find
\begin{equation}
	\delta_{1} - \delta_2 = \delta_{3g} - \delta_3 = \frac{1}{2}\left(\delta_{4g} - \delta_3 \right) = \delta_{1c} - \delta_{2c} = -\frac{C_2(G) g^2(3+\xi_A)}{64 \pi^2 \epsilon },
\end{equation}
indicating that gravitational interaction does not compromise gauge symmetry.

\newpage

\begin{figure}[h]
	\begin{center}
		\includegraphics[angle=0 ,width=14cm]{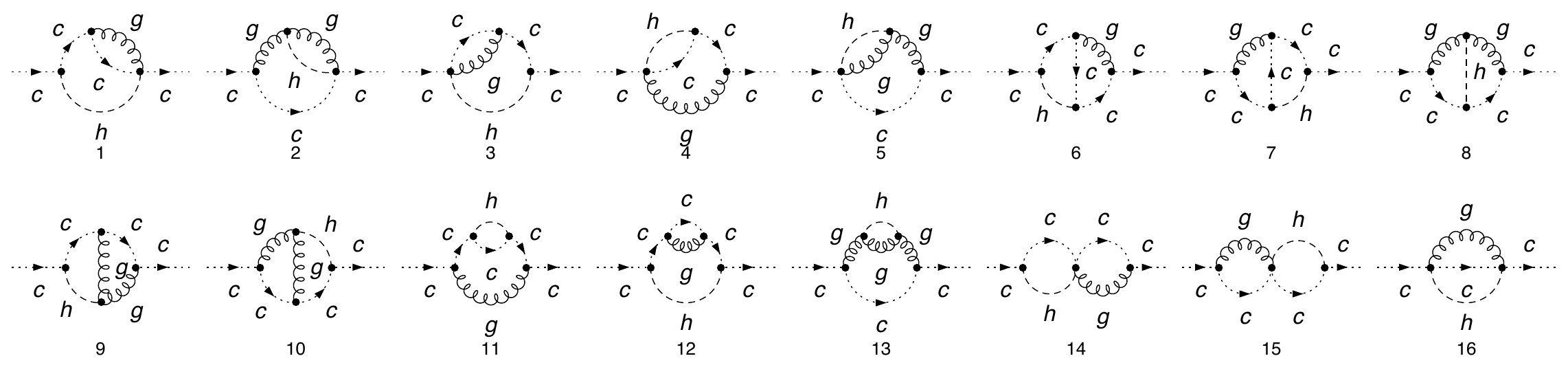}
		\caption{Feynman diagrams illustrating gravitational corrections to the gauge ghost self-energy. Curly, dashed, and pointed lines denote the gluon, graviton, and gauge ghost propagators, respectively. It is noteworthy that these diagrams do not include matter loops, resulting in contributions proportional to $\kappa^2 p^4$. }
		\label{figGhostSE01}
	\end{center}
\end{figure}

\begin{figure}[h]
	\begin{center}
	\includegraphics[angle=0 ,width=14cm]{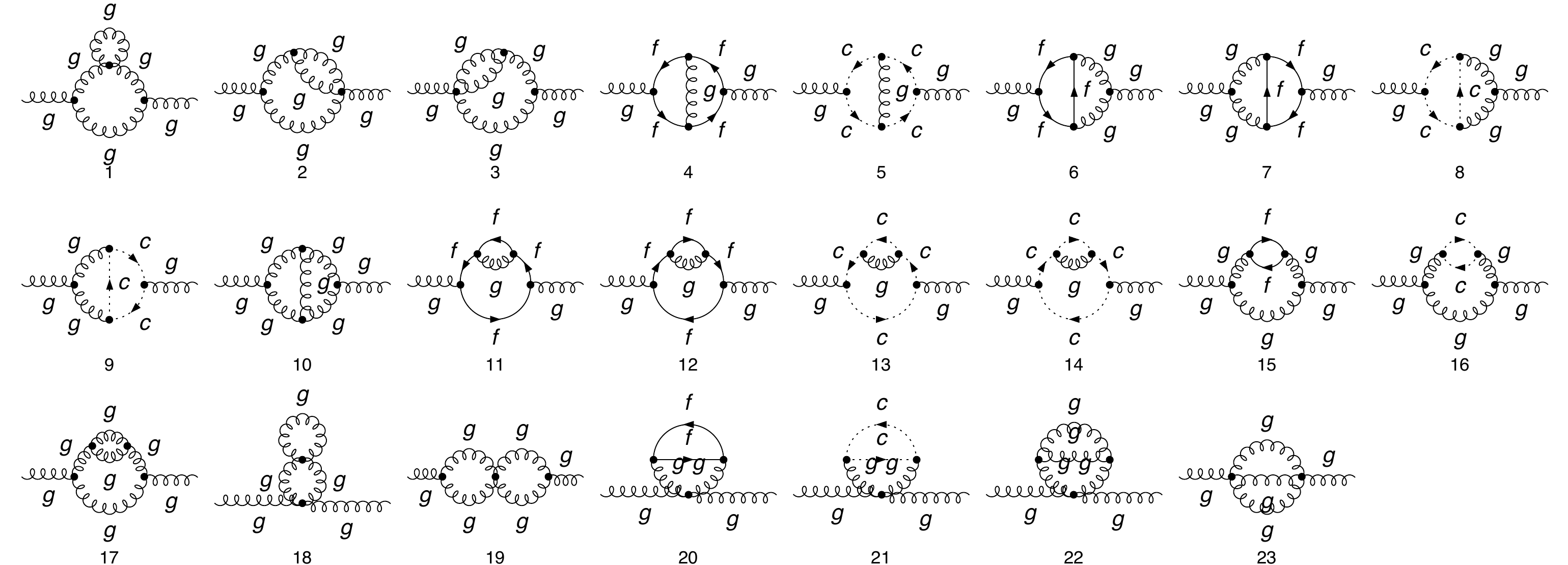}
	\caption{Feynman diagrams depicting the gluon self-energy. Curly and straight lines symbolize the gluon and fermion propagators, respectively.}
	\label{figSE01}
	\end{center}
\end{figure}

\begin{figure}[h!]
	\begin{center}
	\includegraphics[angle=0 ,width=14.5cm]{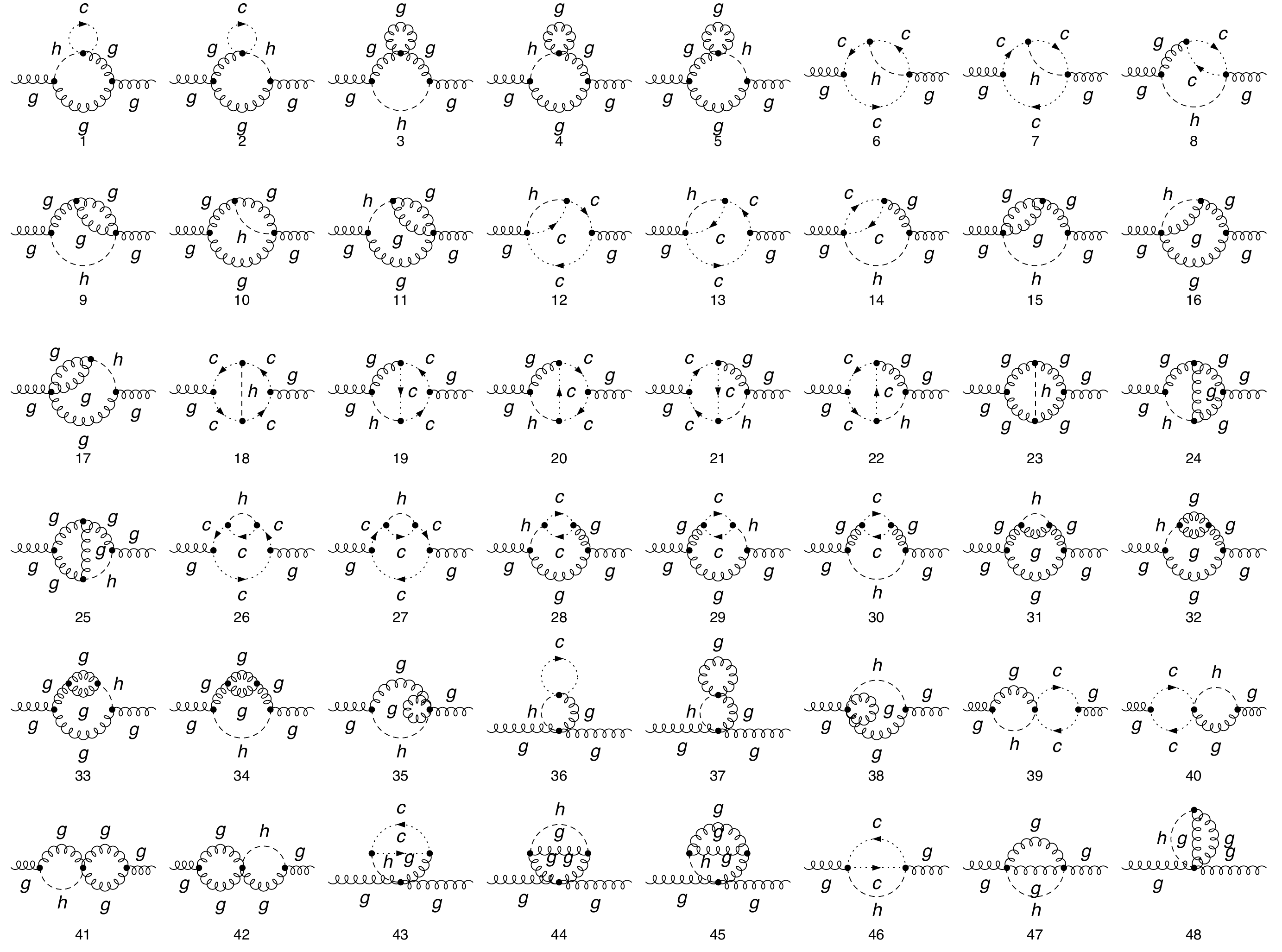}
	\caption{Feynman diagrams illustrating the gluon self-energy incorporating gravitational interaction. These diagrams consist of terms proportional solely to $\mathcal{O}(\kappa^2 p^4)$.}
	\label{figSE02}
	\end{center}
\end{figure}

\begin{figure}[h!]
	\centering
	\includegraphics[angle=0 ,width=16cm]{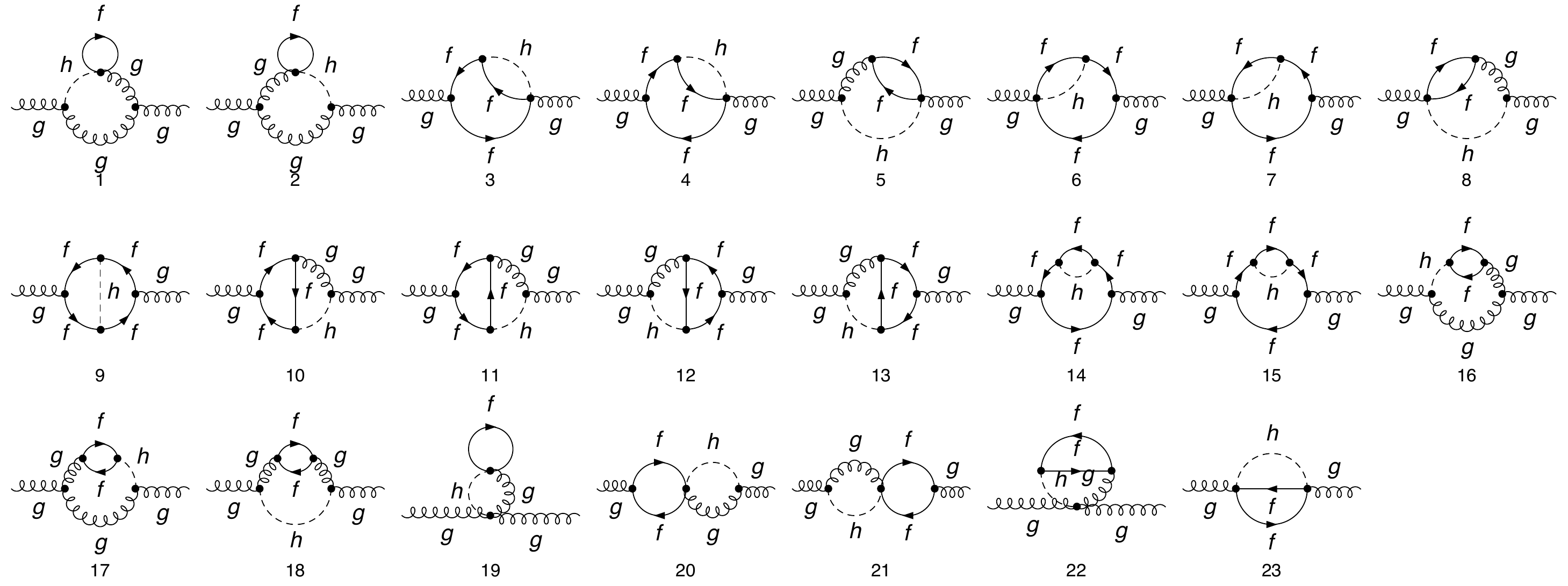}
	\caption{Feynman diagrams depicting the gluon self-energy with contributions from matter and graviton propagators. These diagrams include terms proportional to $\kappa^2 m^2$ for $\Pi(p)$, in addition to $\kappa^2 p^2$.}
	\label{figSE03}
\end{figure}

\begin{figure}[h!]
	\centering
	\includegraphics[angle=0 ,width=16cm]{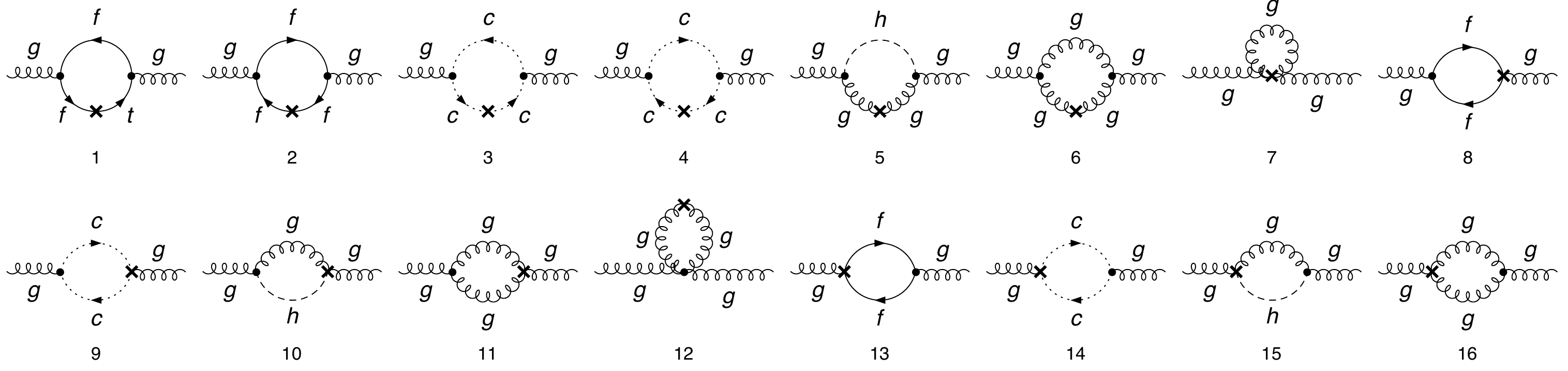}
	\caption{One-loop Feynman diagrams illustrating the gluon self-energy with counterterm insertions. These diagrams are of the same order as the two-loop diagrams.}
	\label{figSECT}
\end{figure}

\end{document}